\documentclass[twoside,slac_one]{revtex4}
\usepackage{graphicx,epsfig}
\usepackage{fancyhdr}
\usepackage{amsmath} 
\usepackage{bm}
\usepackage{amsxtra}
\usepackage{amssymb}
\usepackage{amsthm}
\usepackage{latexsym}
\usepackage{lscape}

\begin{document}

\title{The Fermion Sign Problem at Finite Density, \\
and Large $N_c$ Orbifold Equivalence}

%

\author{A.~Cherman}
\email{a.cherman@damtp.cam.ac.uk}
\affiliation{Department of Applied Mathematics and Theoretical Physics, Cambridge University,  CB3 0WA, UK}

\author{B.~C.~Tiburzi}
\email{btiburzi@ccny.cuny.edu}
\affiliation{Center for Theoretical Physics,
        Massachusetts Institute of Technology,
        Cambridge, MA 02139, USA}
\affiliation{      
        Department of Physics,
        The City College of New York,  
        New York, NY 10031, USA}
\affiliation{        
        Graduate School and University Center,
        City University of New York,  
        New York, NY 10016, USA}
\affiliation{        
        RIKEN BNL Research Center, 
        Brookhaven National Laboratory, 
        Upton, NY 11973, USA}

\begin{abstract}
The study of QCD at finite baryon density is severely hampered by the so-called fermion sign problem. 
As a result, 
we have no known first principles approach to study nuclear matter, 
or neutron stars from QCD. 
On the surface, 
the large 
$N_c$ 
limit does not seem to simplify matters. 
In this limit, 
however, 
one can exploit dualities that exist between strongly coupled gauge theories. 
Our focus will be on some rather novel orbifold equivalences that have recently been discovered at finite density. 
These equivalences relate strongly coupled theories plagued by a sign problem, 
to strongly coupled theories free of sign problems. 
As a result, 
such dualities give deeper insight into the nature of the sign problem and possibly provide a way to simulate QCD at finite density in the large 
$N_c$ 
limit.
\end{abstract}

\maketitle

\thispagestyle{fancy}


\section{Motivation and Scope}

Lattice gauge theory provides a quantitative, 
numerical technique for answering non-perturbative questions about strongly interacting theories from first principles. 
As simulations of QCD and other gauge theories are carried out in Euclidean space by computational necessity, 
there is a fundamental restriction on the types of non-perturbative questions that can be answered using lattice techniques. 
In particular, 
the study of QCD at finite baryon density is not possible using Euclidean lattice simulations.
This is due to the fermion sign problem.
In the statistical evaluation of the QCD path integral,
one uses the weight
\begin{equation}
\mathcal{P}[A_\mu]
= 
\text{Det} 
\left( \mathcal{D}[A_\mu] \right)
e^{ - S_{\text{YM}}[A_\mu]}
,\end{equation} 
to generate an ensemble of gauge configurations, 
$A_\mu(x)$,
in order to sample the important contributions to the QCD partition function. 
Notice we employ a continuum formulation throughout for simplicity. 
The determinant factor arises from integrating out fermionic fluctuations, 
and depends on the fermion action, 
$\mathcal{D}$, 
which has the form
\begin{equation}
\mathcal{D}[A_\mu]
=
(\partial_\mu + i A_\mu ) \gamma_\mu + \mu \gamma_4 + m  
.\end{equation}
Here we have included a chemical potential coupled to fermion number, 
$\mu$.\footnote{ 
In QCD, the quark chemical potential, 
$\mu_q$, 
is trivially related to the baryon chemical potential, 
$\mu_B$,
by the relation
$\mu_B = 3 \mu_q$.  
We will often be sloppy and simply refer to the quark chemical potential as the baryon chemical potential.
One should keep in mind that in 
$SU(N_c)$ 
gauge theory, 
there is a factor of 
$N_c$ 
relating the two. 
}
At vanishing chemical potential, 
$\mu = 0$,
the eigenvalues of 
$\mathcal{D}[A_\mu]$
come in complex conjugate pairs. 
The determinant is consequently real, 
and one can use 
$\mathcal{P}[A_\mu]$
as a probabilistic weight to generate gauge configurations. 
On the other hand, 
when 
$\mu \neq 0$, 
the fermion sign problem manifests itself as a complex-action problem:
$\mathcal{P}[A_\mu]$
is complex-valued, 
and thus cannot be utilized as a probabilistic weight.

As a consequence of the fermion sign problem, 
we have no known first principles technique to address the physics of QCD at finite baryon density. 
For systems with baryon number on the order of a few, 
one can build these light nuclei three-quarks-at-a-time using lattice QCD correlation functions. 
While there is a signal-to-noise problem infecting such computations, 
impressive progress is being made.
Current simulations are now reaching baryon numbers of
$N_B = 3$ and $4$ for the first time, 
see~\cite{Beane:2009gs,Yamazaki:2009ua}.
Beyond light nuclei, 
one wishes to understand how QCD degrees of freedom manifest themselves in larger nuclei
(such as observed in deep inelastic scattering off nuclei, the so-called EMC effect), 
and in nuclear matter
(such as in the passage of jets through dense media in heavy ion collisions). 
Finally, 
the study of dense quark matter is likely relevant to describe the interior of compact stars. 
The density in the interior of a neutron star, 
for example, 
might be great enough to support Cooper pairing of quarks leading to color superconductivity, 
see~\cite{Rajagopal:2000wf,Alford:2007xm}. 
Determining whether this is actually the case remains an outstanding theoretical challenge. 
While models used to study high density QCD are becoming more realistic, 
a definitive answer directly from QCD remains elusive.

Unfortunately,
we shall not attempt to solve the QCD sign problem here. 
Instead, 
we shall garner insight into the problem by studying certain QCD-like theories. 
A QCD-like theory we take to be any gauge theory with fermion degrees of freedom, 
which does not suffer a sign problem at finite density. 
For example, 
it is well known that 
$SU(2)$ 
gauge theory does not suffer from the fermion sign problem at finite baryon density, 
and has been the subject of numerous theoretical and lattice studies;
see, 
for example, 
two early works on the subject,~\cite{Kogut:1999iv,Hands:1999md}. 
There are various other gauge theories which do not suffer from a sign problem at finite density. 
For gauge groups of arbitrary rank,
there is also: 
$SU(N_c)$ 
gauge theory at finite isospin density, 
$SU(N_c)$
gauge theory with adjoint matter at finite baryon density, 
$SO( N_c)$ 
gauge theory at finite baryon density, 
and 
$Sp( 2 N_c)$
gauge theory at finite baryon density
(the latter two can also be considered at finite isospin density). 
While these theories are each amenable to Euclidean space Monte-Carlo, 
a natural question emerges:
how like QCD are these QCD-like theories?
As we shall focus on the large 
$N_c$ 
limit of QCD,
we wish to know specifically how these theories are like 
$SU(N_c)$ 
gauge theory at finite baryon chemical potential.%
\footnote{
The phenomenological relevance of learning about the behavior of large 
$N_c$ 
QCD depends on the closeness of the large 
$N_c$ 
world to ours, 
which is a subtle question discussed, 
for example, 
in~\cite{Shuster:1999tn,Park:1999bz,Frandsen:2005mb,McLerran:2007qj,Buchoff:2009za,Torrieri:2010gz}.
}

Surprisingly it can be shown precisely how these theories are related in the large 
$N_c$ 
limit,~\cite{Hanada:2011ju}. 
The quantitative relation between these QCD-like theories can be elucidated due to recently discovered orbifold equivalences at finite density. 
The first such equivalence was found in~\cite{Cherman:2010jj}. 
Instead of exploring the universality of QCD-like theories at finite density, 
we shall focus on just one QCD-like theory, 
namely that of 
$SO(2 N_c)$ 
gauge theory at finite baryon chemical potential. 
This theory has yet to be studied using lattice Monte-Carlo techniques, 
but could provide quantitative information about QCD at finite density. 
Our focus throughout is on this connection. 
We begin by exploring the orbifold equivalence of 
$SO(2 N_c)$ 
gauge theory to 
$SU(N_c)$ 
gauge theory at finite baryon density. 
While the orbifold equivalence holds in perturbation theory, 
it has been shown to break down non-perturbatively as the baryon chemical potential is increased. 
We then investigate whether certain deformations of  
$SO( 2 N_c)$
gauge theory can extend the validity of the equivalence to larger values of the baryon chemical potential,%
~\cite{Cherman:2010jj,Cherman:2011mh}.
Our presentation ends with a brief summary, 
and an outlook to future work.

\section{Orbifold Equivalence at Finite Density}

The gauge group
$SO(2N_c)$
contains
$SU(N_c)$ 
as a subgroup. 
Projecting out everything but these 
$SU(N_c)$
degrees of freedom from 
$SO(2 N_c)$
allows us to establish an orbifold equivalence between the two theories valid in the large 
$N_c$ 
limit.  
As we wish to include matter fields as well, 
we will spell out the case of a more general orbifold equivalence between two theories. 
The notion of orbifold equivalence was first discovered in the context of string theory,~\cite{Kachru:1998ys}. 
Shortly after it was realized that such equivalences also applied to supersymmetric 
gauge theories, as well as non-supersymmetric theories, see~\cite{Bershadsky:1998cb,Schmaltz:1998bg}.

For our purposes, 
we spell out an orbifold equivalence in three steps, 
and specialize to the particular case of interest below.

\begin{enumerate}
\item
Consider a gauge theory, which we shall refer to as the parent theory  
$(\mathbb{P})$.
Suppose that 
$\mathbb{P}$ 
is invariant under a discrete subgroup 
$\mathbb{Z}_\Gamma$.  

\item
Form what is referred to as the daughter gauge theory 
$(\mathbb{D})$,
which consists of all 
$\mathbb{Z}_\Gamma$-invariant 
fields of the parent theory
$\mathbb{P}$.

\item
There is a planar equivalence between correlation functions of 
$\mathbb{Z}_\Gamma$-neutral 
fields of 
$\mathbb{P}$ 
and the corresponding correlators in 
$\mathbb{D}$. 

\end{enumerate}

\noindent
It is useful to spell this equivalence out symbolically. 
Writing the fields of the parent gauge theory
$\mathbb{P}$ 
as 
$\phi_\mathbb{P}^{Q_\Gamma}$, 
where 
$Q_\Gamma$ 
is the charge of the field 
$\phi_{\mathbb{P}}$
under
$\mathbb{Z}_\Gamma$, 
and the fields of the daughter gauge theory 
$\mathbb{D}$ 
as 
$\varphi_\mathbb{D}$, 
we can write the orbifold equivalence of correlators generically in the form
\begin{equation}
\langle \cdots \phi_{\mathbb{P}}^{Q_\Gamma = 0} \cdots \rangle
\overset{N_c \to \infty}{=}
\langle \cdots \varphi_{\mathbb{D}} \cdots \rangle
.\end{equation}
The equivalence generally relates only a subset of the correlation functions that exist in 
$\mathbb{D}$.

The idea of using a orbifold equivalence at finite density is to embed the sign problem in the equivalence. 
Thus one hopes to have an equivalence between a theory without a sign problem, 
and one with a sign problem. 
In the non-perturbative regime, 
both theories in the equivalence are strongly coupled;
however, 
the parent is amenable to lattice Monte-Carlo, 
while the daughter is not. 
Hence, 
one can use the equivalence to compute certain correlators of the daughter theory indirectly from the parent.

Now consider 
$SO(2 N_c)$
gauge theory coupled to matter in the fundamental representation. 
At any value of the baryon chemical potential, 
the theory has the following discrete symmetry
\begin{equation} \label{eq:Z2}
\mathbb{Z}_2 :
\quad
\begin{pmatrix}
0 & 1_{N_c} \\
- 1_{N_c} & 0
\end{pmatrix}
\otimes 
e^{ i \pi / 2}
\in SO(2N_c) \otimes U(1)_B
.\end{equation}
The daughter theory formed from 
$\mathbb{Z}_2$-invariant 
operators can be shown to be
$SU(N_c)$ 
gauge theory%
\footnote{
Strictly speaking, 
the daughter theory actually has the gauge group 
$U(N_c)$; 
however, 
the difference between this gauge group and 
$SU(N_c)$
is suppressed at large $N_c$. 
} 
with fundamental matter and a baryon chemical potential,~\cite{Cherman:2010jj}. 
Thus one can compute correlations functions of neutral operators in 
$SO(2N_c)$ 
gauge theory to learn about large 
$N_c$ 
QCD at finite baryon density. 
One such correlation function, 
for example, 
is the chiral condensate,
$\langle \overline{\psi} \psi \rangle$.
A crucial point to note is that the discrete symmetry must involve the baryon number transformation in order to project onto 
$SU(N_c)$ 
gauge theory at finite baryon density.

\section{Equivalence Breakdown}

We have spelled out the orbifold equivalence between a sign-problem-free theory, 
$SO( 2 N_c)$ at finite chemical potential, 
and a sign-problematic theory, 
large $N_c$ QCD at finite chemical potential.
At this point, 
it just seems we are hiding the fermion sign problem in a complicated theoretical apparatus. 
In fact, 
this is not too far from the truth. 
While orbifold equivalences hold to all orders in perturbation theory, 
there are well-known problems with such equivalences at the non-perturbative level, 
see,~\cite{Tong:2002vp,Armoni:2005wta,Kovtun:2005kh,Unsal:2008ch}. 
Generally what happens is that phase transitions can spoil orbifold equivalences. 
For example, 
the parent theory can develop condensation which spontaneously breaks the symmetry involved in the orbifold projection. 
A heuristic picture of why such symmetry breaking is detrimental to orbifold equivalences is sketched in~\cite{Cherman:2011mh}.

For the specific orbifold equivalence at hand, 
we must realize that baryon number can be spontaneously broken in theories with an orthogonal gauge group. 
This is because such theories contain two types of baryons, 
the type-$A$ baryons formed of 
$N_c$
fermions, 
and the type-$B$ baryons formed of just two fermions, 
\begin{eqnarray}
\mathcal{B}_A 
&\sim& 
\epsilon_{ijk \cdots} \, \psi_i \psi_j \psi_k \cdots
\notag \\
\mathcal{B}_B
&\sim&
\psi^T_i \psi_i
\notag
.\end{eqnarray}
The diquark interpolating field, 
$\mathcal{B}_B$,
while gauge invariant, 
is not invariant under the 
$\mathbb{Z}_2$
symmetry used in the orbifold projection, 
Eq.~\eqref{eq:Z2}, 
because
$\mathcal{B}_B \overset{\mathbb{Z}_2}{\to} - \mathcal{B}_B$. 
In terms of correlators, 
we might not be too worried about such states since they have no image in the daughter theory. 
Unfortunately this is na\"ive.  
Because the $\mathcal{B}_B$ are charged under baryon number,  
the chemical potential will lower the energy of such states. 
Once the chemical potential is large enough, 
it will become energetically favorable for them to condense. 
The  
$\mathbb{Z}_2$
symmetry will then be spontaneously broken. 
One might further hope that the size of the chemical potential needed for such condensation is large;
however, 
it turns out that these states, 
having scalar diquark quantum numbers, 
are abnormally light.
The 
$SO( 2 N_c)$
gauge theory with 
$N_f$
light quarks has an enhanced 
$SU(2 N_f)$ 
flavor symmetry%
\footnote{
At large $N_c$, 
the effect of the chiral anomaly vanishes, 
giving one a
$U(2 N_f)$ symmetry. 
Consequently there is one more Goldstone mode, 
which corresponds to the flavor-singlet 
$\eta^\prime$ 
meson. 
} 
that is spontaneously broken down to 
$SO(2 N_f)$ 
at vanishing chemical potential by the formation of the chiral condensate,~\cite{Coleman:1980mx}. 
The resulting 
$N_f ( 2 N_f + 1) - 1$ 
Goldstone modes consist of 
$N_f^2 - 1$ 
pseudoscalar pions, 
and 
$N_f ( N_f + 1)$
baryonic pions,~\cite{Peskin:1980gc,Cherman:2010jj,Cherman:2011mh}.
The latter turn out to have the quantum numbers of 
$\mathcal{B}_B$
and 
$\overline{ \mathcal{B}}_B$. 
The 
$\mathcal{B}_B$
modes will condense at zero temperature for a rather small value of the chemical potential,
$\mu_B = \frac{1}{2} m_\pi$,
and invalidate the equivalence. 
Thus the equivalence is valid for 
$\mu_B < \frac{1}{2} m_\pi$.

\begin{figure}
\includegraphics[width=0.5\columnwidth]{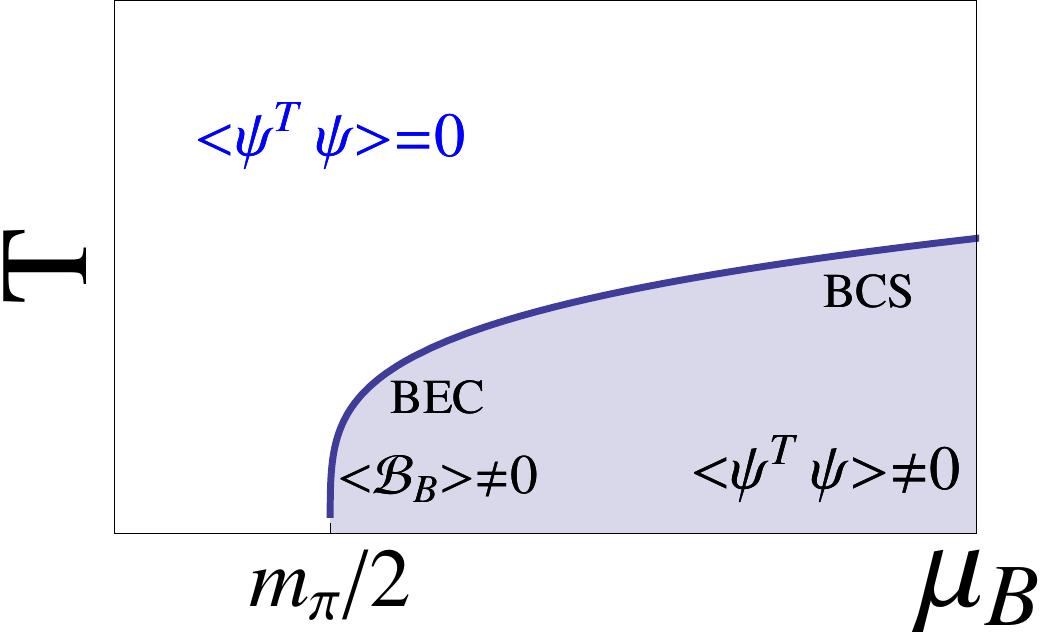}%
\caption{\label{f:SOphase}
Phase diagram of 
$SO(2 N_c)$ 
gauge theory.
The region of spontaneously broken baryon number corresponds to the region where orbifold equivalence to large 
$N_c$ 
QCD breaks down.
Outside this shaded region, 
the equivalence is valid. 
Other phase transitions (chiral, deconfinement) are not shown.}
\end{figure}

The phase diagram for 
$SO(2 N_c)$
gauge theory is shown in Fig.~\ref{f:SOphase}, 
where we depict the region of spontaneous baryon number violation. 
For moderate values of the chemical potential, 
a Bose-Einstein condensate (BEC) of baryonic-pions will form. 
At asymptotically high densities, 
one can show that Cooper pairs will form in the scalar diquark channel. 
These two regions are very likely continuously connected.%
\footnote{
What is remarkable 
(and true of all QCD-like theories), 
is that in the asymptotic regime while BCS pairing is supported, 
there is no superconductor, 
because the BCS pairs are gauge invariant. 
QCD appears to be special, 
because BCS pairing is not gauge invariant.
} 
Notice that the phase diagram for 
$SO(2 N_c)$ 
gauge theory with baryon chemical potential is qualitatively similar to 
$SU(N_c)$
gauge theory at finite 
\emph{isospin} 
density, 
which was first studied in~\cite{Son:2000xc}. 
In fact, 
the similarities can be made quantitative in the large $N_c$ limit, 
because there is an orbifold equivalence relating the two theories,~\cite{Hanada:2011ju}.
This orbifold equivalence does not involve a baryon number transformation, 
and is conjectured to be valid in the entire $\mu$-$T$ plane.

\section{Deformations}

We have seen that the orbifold equivalence between 
$SO(2N_c)$
gauge theory at finite baryon density and large 
$N_c$ 
QCD at finite baryon density only holds in a limited region. 
Beyond a certain value of the chemical potential 
(which may be a function of temperature), 
the equivalence becomes invalid due to spontaneous breaking of 
$U(1)_B$. 
Our goal is now to extend the reach of the equivalence by considering deformations of the parent theory. 
Such deformations are efficaciously designed to obstruct transitions to the broken phase.

Our motivation to consider such deformations is inspired by recent progress in a related matter, 
namely Eguchi-Kawai reduction and the volume independence of Yang-Mills at large 
$N_c$, 
see~\cite{Eguchi:1982nm,Bhanot:1982sh,GonzalezArroyo:1982hz}, 
which can be rephrased in the language of an orbifold equivalence,
\cite{Kovtun:2007py}. 
Large 
$N_c$ 
volume independence relates gauge theories in different physical volumes provided that they remain in their center-symmetric phases. 
To study Yang-Mills on the lattice, 
it would be ideal to reduce the theory to the smallest possible volume: 
a single plaquette. 
Under this extreme volume reduction, 
center symmetry is spontaneously broken, 
and one can no longer study the physics of the confined phase. 
To rescue volume independence, 
it has been argued,~\cite{Unsal:2008ch},
that one can add certain double-trace terms to the original Yang-Mills action. 
The resulting deformed theory is the same as Yang-Mills in the large $N_c$ limit. 
The double trace terms, 
however, 
can be chosen to prevent the breaking of center symmetry, 
and enable the volume reduction to a single plaquette. 
This idea is being tested in lattice simulations, 
and has been met with favorable success,~\cite{Bringoltz:2009kb}.

For the case at hand, 
we wish to stabilize $SO(2 N_c)$ gauge theory at finite density  against the breaking of 
$U(1)_B$.
Baryon number breaks spontaneously as the chemical potential is increased because the baryonic pion energy is lowered to zero. 
The purpose of the deformation in this case is to lift the mass of the baryonic pion delaying the onset of the phase transition. 
If the baryonic pions were elementary fields of the theory, 
we could simply add a mass term for these modes. 
In terms of elementary fermion fields, 
however, 
the best we can do is add a four-fermion term that leads to a repulsive interaction between fermions coupled in the scalar diquark channel. 
Two particularly useful choices for the deformation are given explicitly by,~\cite{Cherman:2010jj,Cherman:2011mh},
\begin{equation}
V_{\pm}
= 
C^2
\sum_{a,b}^{N_f}
\left(
S^\dagger_{ab} S_{ab}
\pm 
P^\dagger_{ab} P_{ab}
\right)
,\end{equation}
where 
$S_{ab} = \psi_a^T C \gamma_5 \psi_b$ 
has the quantum numbers of a scalar baryonic pion, 
and 
$P_{ab} = \psi^T_a C \psi_b$
has the quantum numbers of a pseudo-scalar baryonic meson. 
Because 
$C^2 > 0$,
scalar diquark scattering is repulsive, 
and the mass of the
$\mathcal{B}_B$ 
baryonic pions should be additively renormalized by an amount proportional to 
$C^2$.
If this is indeed the case, 
the 
$U(1)_B$
breaking phase transition has been offset by an amount proportional to 
$C^2$.

Although the deformation is physically well-motivated, 
the non-perturbative dynamics of the system determine its effects.  
In general, the study of the deformed theory requires a lattice computation. 
One can use effective field theory techniques, 
however,  
to investigate the effects of the deformation at low-energies. 
When the fermion mass, 
chemical potential, 
and deformation are set to zero, 
the low-energy dynamics are described in terms of the emergent Goldstone modes.  
These modes are packaged in the coset field, 
$\Sigma \in SU(2 N_f) / SO (2 N_f)$. 
One then treats the chemical potential and fermion mass terms as perturbations about the chiral limit. 
The same can be done for the four-fermion operators contained in the deformation. 
In fact, 
four-fermion operators have long been treated in chiral perturbation theory, 
see~\cite{Bernard:1985wf}. 
A particularly simple way to account for such four-fermion operators is to use spurions, 
for a related example, 
see~\cite{Bar:2003mh,Tiburzi:2005vy}.
Using a spurion analysis, 
the effects of the deformation on the low-energy theory have been determined,~\cite{Cherman:2011mh}. 
For example, 
chiral perturbation theory for the 
$V_+$
deformed theory contains an additional term, 
\begin{equation}
V_+^{\text{eff}} 
=
v_+
\sum_{a,b}^{N_f}
\left[
\text{Tr} 
\left(
\Sigma L^{(ab)}
\right)
\text{Tr} 
\left(
\Sigma^\dagger
L^{(ab)\dagger} 
\right)
+
\text{Tr} 
\left(
\Sigma R^{(ab)}
\right)
\text{Tr} 
\left(
\Sigma^\dagger
R^{(ab)\dagger} 
\right)
\right]
,\end{equation}
where the 
$L^{(ab)}$ 
and 
$R^{(ab)}$ 
are the spurion fields used to make the interaction symmetric under 
$SU(2 N_f)$.
At the end of the day, 
these fields take on prescribed constant values. 
The parameter
$v_+$
is a new low-energy constant. 
Its value is directly proportional to 
$C^2$, 
and has been shown to be positive.

%
\begin{figure}
\includegraphics[width=0.45\columnwidth]{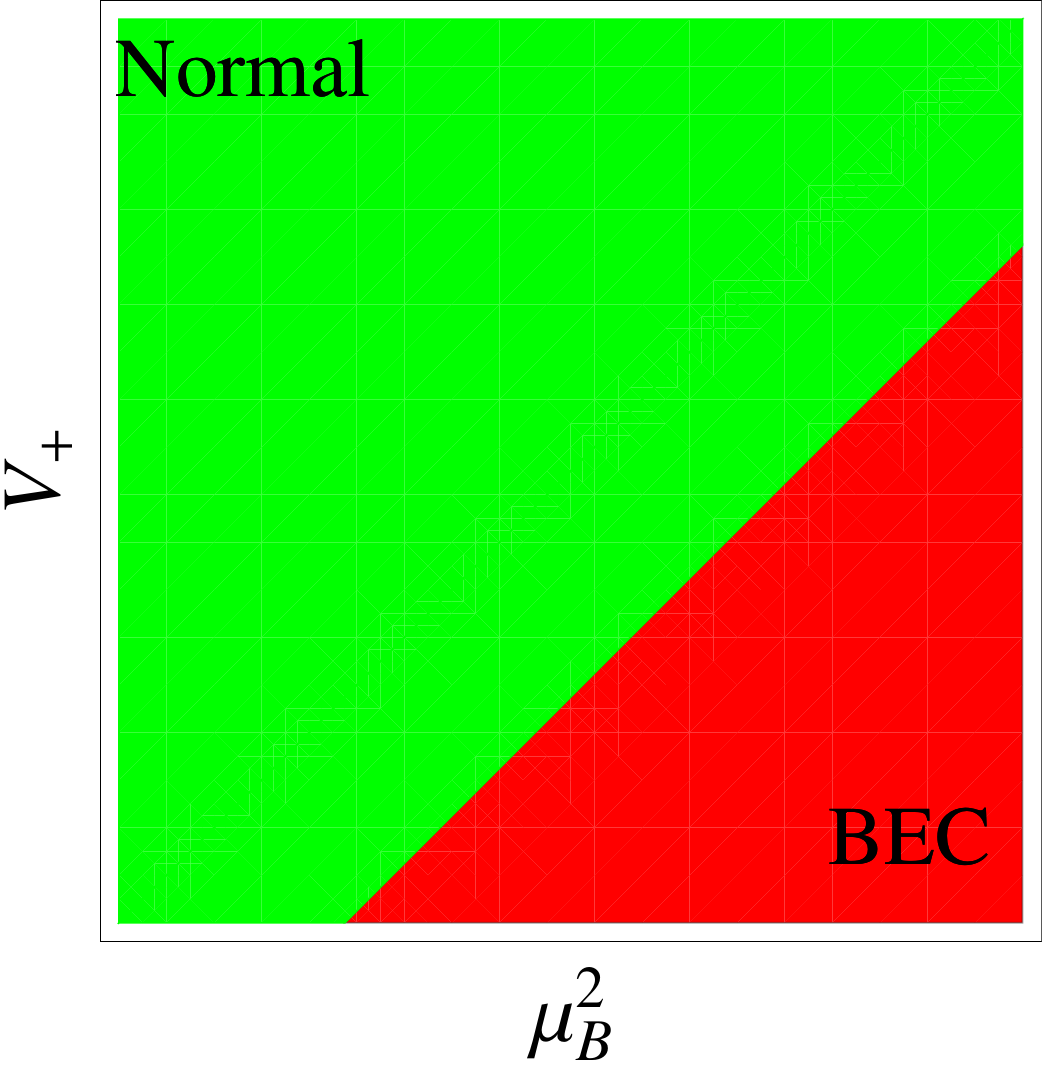}%
$\qquad$
\includegraphics[width=0.45\columnwidth]{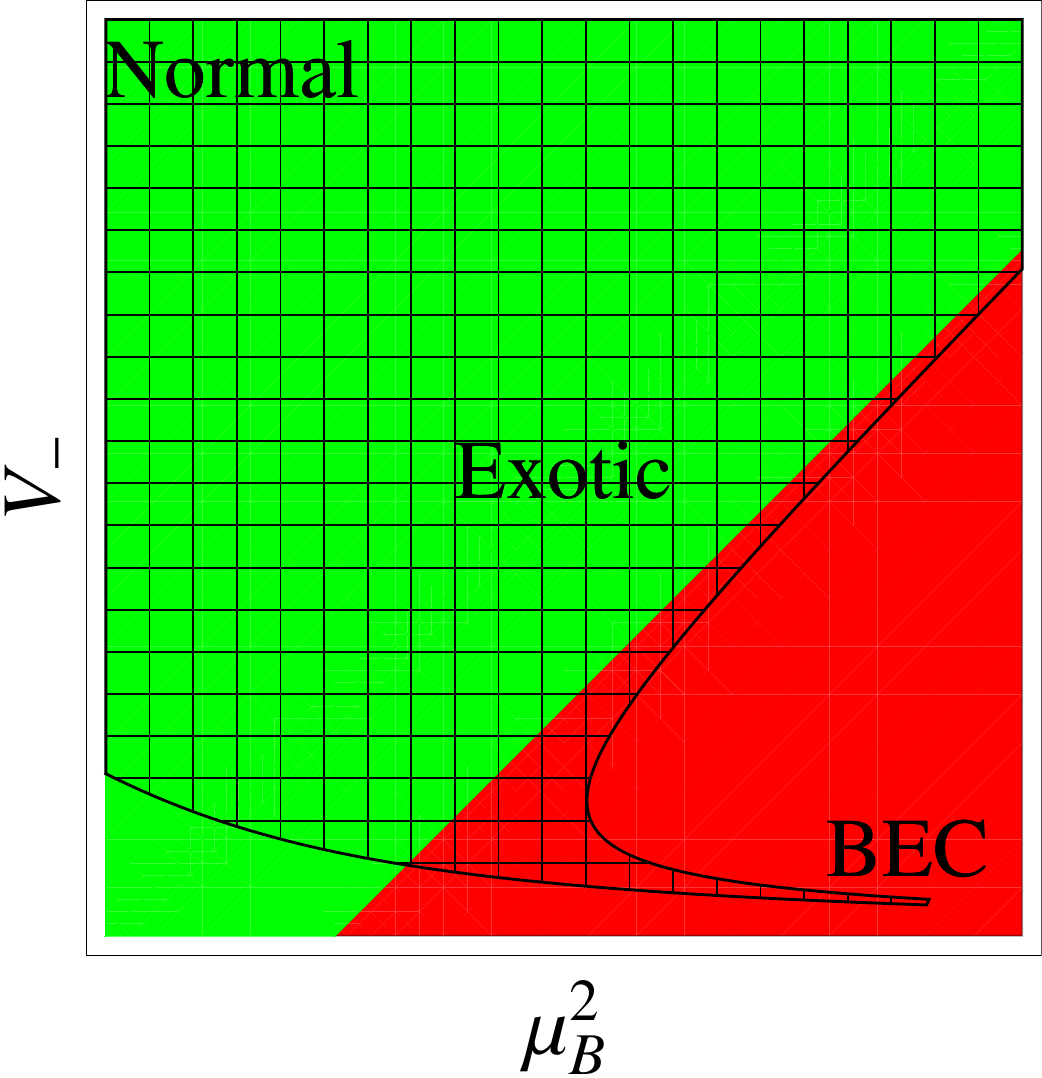}%
\caption{\label{f:Deform}
Schematic phase diagram of deformed
$SO(2 N_c)$ 
gauge theory.
The $x$-axis, 
where 
$V_\pm = 0$, 
corresponds to the undeformed theory, 
and
$V_\pm$ 
represents the strength of the deformation. 
Regions where baryonic-pions form a BEC correspond to those regions where the orbifold equivalence to large 
$N_c$ 
QCD breaks down.
Notice that one can remain outside the BEC phase for a large enough value of the deformation. 
There is a large $N_c$ exotic phase characterized by a BEC of baryonic pions in addition to an $\eta^\prime$ condensate. 
Such a phase only exists in the $V_-$-deformed theory, and has been shown to be metastable.    
}
\end{figure}
%

Expanding the 
$\Sigma$ 
field out in terms of Goldstone modes, 
$\Sigma = 1 + i \Phi + \ldots$, 
the above operator gives rises to masses of the 
$\mathcal{B}_B$ 
modes. 
These masses are proportional to 
$v_+$
and so delay the onset of the
$U(1)_B$  
breaking phase transition. 
In fact, 
the vacuum structure of the 
$V_\pm$-deformed 
theories has been analyzed in detail using the effective theory.
In Fig.~\ref{f:Deform}, 
we show the schematic phase diagrams of the deformed theories as a function of the baryon chemical potential squared. 
From the figure, 
one sees that the BEC phase can be avoided by tuning the deformation to be large enough. 
As a result, 
one can remain in the normal phase for which the orbifold equivalence is valid. 
The effective theory, hence, points to success for the deformations, 
however, 
there are a number of practical issues that must be resolved in order to study the deformed theories using lattice gauge theory techniques. 
We mention some of these issues in our concluding words.

\section{Summary and Outlook}

Above we have investigated the orbifold equivalence between 
$SO(2 N_c)$ 
gauge theory at finite baryon chemical potential 
and 
$SU(N_c)$
gauge theory at finite baryon chemical potential. 
This orbifold equivalence is novel, 
because the former theory does not have a sign problem at finite baryon density, 
while the latter does have a sign problem. 
In regimes where the orbifold equivalence is valid, 
one can study aspects of the sign-problematic theory by using lattice Monte-Carlo simulations
of the sign-problem free theory. 
The correlation functions which have non-trivial images in the sign-problematic theory are identical to those in the sign-problem-free theory up to 
corrections suppressed by powers of $1/N_c$.

Using the technique of orbifold equivalence, 
the fermion sign problem at large 
$N_c$ 
has thus been mapped into a validity problem at the non-perturbative level. 
At relatively small values of the chemical potential, 
$\mu_B \geq \frac{1}{2} m_\pi$, 
the equivalence is known to be invalid due to spontaneous baryon number violation in the parent theory. 
This situation is analogous to that of the phase quenched approximation to QCD at finite baryon density.%
\footnote{
Recall that phase quenched QCD is identical to QCD with an isospin chemical potential. 
}
For small chemical potential, 
it might be harmless to quench the phase of the fermion determinant, 
but at 
$\mu_B \geq \frac{1}{2} m_\pi$, 
the approximation is horrible due to the formation of a pion condensate, 
which has no analogue in the phase unquenched theory. 
Orbifold equivalence shows that all three theories are basically the same at large 
$N_c$
as long as
$\mu_B < \frac{1}{2} m_\pi$. 
Beyond this value, 
only the phase quenched theory and 
$SO(2 N_c)$ 
theory remain equivalent.

To extend the equivalence to larger values of the chemical potential, 
it is necessary to deform the parent theory by terms that are suppressed by powers of 
$1/N_c$. 
Such deformations are designed to preserve 
$U(1)_B$ 
symmetry as the baryon chemical potential is increased.
The effects of the deformation on the dynamics of 
$SO(2 N_c)$, 
however,
can only be determined non-perturbatively. 
Fortunately at low-energies, 
we can investigate the effect of the deformation using effective field theory, 
and the analysis indeed points to the preservation of 
$U(1)_B$.
Our effective field theory analysis assumes the power counting:
$\mu_B^2 \sim m_\pi^2 \sim C^2$.
In this way, 
the four-fermion operators have been promoted from irrelevance to the same order as other terms, 
with these terms considered perturbatively small compared to the chiral symmetry breaking scale. 
To extend the equivalence to progressively larger values of the chemical potential, 
one will very likely need to increase the value of 
$C^2$.
There is a limit to how far 
$C^2$ 
can be increased in a lattice calculation, 
because the deformed theory formally has no continuum limit.
It may be possible that certain aspects of the large 
$N_c$ 
limit allow 
$C^2$
to be taken arbitrarily large, 
and this issue is under active investigation.

A further issue of practical importance is the inclusion of four-fermion operators in the lattice theory. 
Because fermion fluctuations are integrated out to evaluate the path integral, 
one must first have a Gau{\ss}ian integral for the fermions. 
The only known way to trade four-fermion terms for fermion bilinears is by using auxiliary fields. 
Auxiliary fields are integrated in using the Gau{\ss}ian integral
\begin{equation}
e^{- x^4}
=
\frac{1}{\sqrt{\pi}}
\int_{-\infty}^\infty
dy
\, e^{ - y^2 + 2 i x^2 y }
.\end{equation}
In this toy model example, 
the four-fermion interaction,
$x^4$, 
is traded in for a quadratic interaction vertex, 
$x^2 y$,
with an auxiliary field 
$y$.
If one is unlucky, 
the introduction of auxiliary fields can reintroduce the sign problem. 
Currently it is only known how to introduce the 
$V_-$
deformation without a sign problem in the chiral limit,~\cite{Cherman:2010jj}. 
An obvious avenue for future work is to devise a different implementation of the auxiliary fields
that allows one to simulate away from the chiral limit.

Another important problem is to understand the mapping between baryonic observables between the 
$SO(2N_c)$ 
theory and large 
$N_c$ 
QCD. 
Baryon interpolating operators are 
$\mathbb{Z}_2$-neutral 
when 
$N_c$ 
is even, 
suggesting that baryons may be in the common sector of the two theories; 
but, 
the details of applying the orbifold equivalence to operators involving color-space epsilon tensors have not yet been worked out. 
With any luck, 
the deformed 
$SO( 2 N_c)$ 
gauge theory may support a new phase at finite density in a region before the equivalence breaks down. 
It would be very exciting to see this from lattice studies.

Nonetheless, 
we see that the technique of orbifold equivalence has shed new light onto the very difficult problem of finite density QCD. 
We hope that further activity in this direction can help provide deeper insight into the fermion sign problem, and ultimately 
increase our understanding of the QCD phase diagram.

\begin{acknowledgments}
We are grateful to M.~Hanada and N.~Yamamoto for numerous enlightening discussions. 
Work supported in part by the 
U.S.~Dept.~of Energy, 
Office of Nuclear Physics,
under
Grant No.~DE-FG02-94ER-40818 (B.C.T.), 
and the CUNY Research Foundation (B.C.T.).
\end{acknowledgments}

\bigskip 
\bibliography{tib_bib}




\end{document}